\documentclass[aps,twocolumn,preprintnumbers]{revtex4}
\usepackage[utf8]{inputenc}

\usepackage{graphicx}  
\usepackage{subfigure}
\usepackage{multirow}
\usepackage{natbib}
\usepackage{balance}
\usepackage{flushend}
\usepackage{xcolor}

\usepackage{fancyhdr}
\usepackage{longtable}
\usepackage{parskip}
\usepackage[T1]{fontenc}
\usepackage{dcolumn}   

\usepackage{bm}        
\usepackage{amsfonts}  
\usepackage{amsmath}   
\usepackage{amssymb}   

\newcommand{\pwisein}{\left\{ \begin{array}{ll}}
\newcommand{\pwiseout}{\end{array}\right.}

\begin{document}

\title{Intercomponent vortex interaction and dynamics in a uniform Bose quantum liquid}

\author{Rabeea Kiran$^{1}$, Hiroki Saito$^{4}$, and Hsiang-Hua Jen$^{1,2,3}$}

\affiliation {\it $^{1}$ Institute of Atomic and Molecular Sciences, Academia Sinica, Taipei 10617, Taiwan\\
\it $^{2}$ Department of Physics, National Central University, Taoyuan City, 320317, Taiwan\\
\it $^{3}$ Physics Division, National Center for Theoretical Sciences, Taipei 10617, Taiwan\\
\it $^{4}$ Department of Engineering Science, University of Electro-Communications, Tokyo 182-8585, Japan}

\begin{abstract}  

We investigate the interaction and dynamics of vortices belonging to different components of a binary uniform quantum liquid. Using the coupled extended Gross-Pitaevskii equations  with logarithmic beyond-mean-field corrections, we derive the asymptotic interaction potential between well-separated vortices. The interaction potential retains the asymptotic form $\sim s^{-2}$ as a function of the intervortex distance $s$, but its coefficient depends on the background density through the quantum fluctuation. As a result, the interaction between vortices changes from attractive to repulsive at a critical density. Real-time simulations demonstrate a corresponding reversal of orbital motion of a vortex pair from clockwise to counterclockwise, consistent with their Magnus response.
\end{abstract}

\maketitle 
\section{Introduction}

Bose-Einstein condensates (BECs) of ultracold atoms provide a highly controllable platform through tunable interatomic interactions \cite{stingari, pethick}. Within the mean-field approximation \cite{mf1, mf2}, weakly interacting BECs are well described by the Gross-Pitaevskii (GP) equation. In binary liquid of Bose mixtures, however, the Lee-Huang-Yang (LHY) \cite{lhy} correction can stabilize a droplet against collapse and support a self-bound droplet \cite{petrov1, petrov2, petrov3, d1, d2, d3, d4}. In a two-dimensional (2D) model, the fluctuation-corrected energy exhibits a logarithmic density dependence, leading to a minimum of the energy at a finite density. Experimental observations of droplets in atomic mixtures \cite{droplet, freespace, exp} and dipolar gases \cite{dip1, dip2} have stimulated growing interest in quantum liquids. Their collective and topological excitations, such as vortices \cite{vo1, vo2, vo3}, provide a useful probe of how the equation of state influences superfluid behavior beyond the equilibrium properties of the liquid.

Quantized vortices are particularly sensitive to the density of the surrounding fluid. A vortex combines phase winding with a density depletion, and its structure and dynamics are governed by the energetic cost of this deformation. Theoretical studies have established the existence and stability of 2D vortex droplets \cite{vd1, vd2, vd3, vd4, vd5, vd6, vd7, vd8, vd9, vd10, vd11, vd12, vd13} and demonstrated that background density affects vortex structure and stability in uniform Bose quantum liquids \cite{vd14}. Binary systems introduce an additional degree of freedom because phase winding can reside in either component independently. A vortex in one component therefore perturbs the density of the other, providing a mechanism for interaction between spatially separated vortices belonging to different components.

For two-component condensates coupled through their densities, this mechanism produces an interaction distinct from that between vortices in a single component. Eto et al. \cite{bo4} derived the asymptotic intercomponent vortex interaction, which scales as $\ln(s/\xi)/s^2$, where $s$ denotes the intervortex separation and $\xi$ is a short-distance cutoff. The sign of interaction is determined by the intercomponent coupling and is independent of the vortex circulations. Subsequent studies examined the interaction at shorter separation distances and related vortex trajectories to the balance between the interaction force and the Magnus force \cite{orb1}. In circular confinement, boundary-induced flows were shown to modify the vortex dynamics \cite{bo7}. These studies established a framework relating intervortex interaction to vortex motion.

In a binary quantum liquid, the LHY correction introduces a density dependence into the response that mediates the intercomponent vortex interaction. Changing the background density modifies the relative energetic costs of total-density and relative density deformations. As a result, the density disturbance induced by a vortex in the other component can change character, suggesting that the intercomponent vortex interaction may be controlled through the background density even when the microscopic coupling parameters remain fixed. A central question is whether this mechanism can reverse the interaction force and the resulting vortex motion, and whether a critical density exists at which the leading logarithmic interaction vanishes. 

In this work, we investigate the interaction and dynamics of two initially separated half-quantized vortices belonging to different components of a 2D binary quantum liquid. Using the coupled extended GP equations with the logarithmic LHY correction, we derive the asymptotic intervortex interaction potential and identify a density-dependent coefficient that changes sign with the background density. This leads to a critical density separating attractive and repulsive regimes. Real-time simulations show a corresponding reversal of orbital motion: for two vortices with positive unit circulation, clockwise motion at low density changes to counterclockwise motion at high density. At the predicted critical density, the leading asymptotic interaction vanishes, and no resolvable vortex motion is observed over the simulated time interval. These results demonstrate how the fluctuation-modified equation of state controls the intervortex interactions and vortex dynamics in multicomponent quantum liquids.

The paper is organized as follows:
Section~\ref{sec:model} introduces the model, confinement, and numerical method. Section~\ref{s:intervortex} derives the density-dependent intervortex interaction, identifies the critical density, and compares it with numerical results. We explain the reversal of vortex motion and present their numerical dynamics in Sec.~\ref{s:motion}. Section~\ref{sec:conc} summarizes the results and discusses their implications.
\section{Model and Numerical Framework}
\label{sec:model}
\subsection{Extended Gross-Pitaevskii model}
We consider a 2D homonuclear Bose mixture of atoms having mass $m$, with repulsive intracomponent and attractive intercomponent interactions. The two components are coupled through their densities and are not coherently interconverted. Including the leading 2D quantum-fluctuation contribution, the dynamics is described by the coupled extended GP equations \cite{petrov2},
\begin{widetext}
\begin{align}
i\hbar\frac{\partial\Psi_j}{\partial t}
=\left[
-\frac{\hbar^2}{2m}\nabla^2 + V(\textbf{r})
+\Lambda(n_j-n_k)
+Gn\ln\left(\frac{n}{n_r}\right)
\right]\Psi_j, \quad j\ne k, 
\label{eq:dimensional_egpe1}
\end{align}
\end{widetext}
 where $n_j=|\Psi_j|^2$ is the density of $j$th component, $n=n_1+n_2$ is the total density, and $V(\textbf{r})$ is the external potential. The coefficient $\Lambda=\frac{4\pi\hbar^2}{m\ln(|a_{12}|/a)}>0$ is the modified coupling constant, with $a_{kj}$ being the 2D scattering lengths between component $j$ and $k$, and  $a=a_{11}=a_{22}$. It determines the energetic cost of a density imbalance between the components. The logarithmic term describes the fluctuation-corrected response of total density with a coefficient $G=\frac{\Lambda}{\ln(|a_{12}|/a)}>0$. The reference density $n_r$ in the last term is related to the equilibrium density per component $n_0$ through $n_r=2\sqrt{e}n_0$, where
\begin{equation}
n_0=
\frac{e^{-2\gamma_{\mathrm E}-3/2}}{2\pi}
\frac{\ln(|a_{12}|/a)}{a|a_{12}|},
\label{eq:equilibrium_density}
\end{equation}
with $\gamma_{\mathrm E}$ being the Euler's constant. 

We rescale the position $\bm{r}=\Tilde{\bm{r}}\lambda$, time
$t=\sigma\tau$, and the wave function $\Psi_{j}(\bm{r},t)=\sqrt{n_r}\Tilde{\Psi}_{j}(\Tilde{\bm{r}}, \tau)$, where $\lambda=\hbar/\sqrt{2mGn_r}$ and $\sigma=\hbar/(Gn_r)$. Writing Eq.~\eqref{eq:dimensional_egpe1} in a dimensionless form,
\begin{equation}
  \begin{aligned}
&i\frac{\partial\Tilde{\Psi}_j}{\partial\tau}=
\left[
-\Tilde{\nabla}^2
+\Tilde{V}(\Tilde{\textbf{r}})+2\beta(\Tilde{n}_j-\Tilde{n}_k)
+\Tilde{n}\ln\Tilde{n}
\right]\Tilde{\Psi}_j,
\label{eq:dimensionless_egpe2}
\end{aligned}
\end{equation}
where
\begin{equation}
\beta=\frac{\Lambda}{2G}=\frac{1}{2}\ln\Bigg(\frac{|a_{12}|}{a}\Bigg)
\end{equation}
is the scaled interaction coefficient and $\Tilde{V}=V/(Gn_r)$ is the scaled
potential. We omit the tildes from the
dimensionless quantities below.

The associated dimensionless energy functional is defined as
\begin{equation}
E[\Psi]= E_{\textrm{kin}}[\Psi] + E_{\textrm{int}}[\Psi]+ E_{\textrm{trap}}[\Psi].
\label{eq:dimensional_energy}
\end{equation}
Here, the kinetic, contact interaction, and potential energies are given by,
\begin{equation}
E_{\textrm{kin}}[\Psi]=\int d^2\mathbf{r}\sum_{j=1}^{2}|\nabla\Psi_j|^2, 
\label{eq:kin_energy}
\end{equation}
\begin{equation}
E_{\textrm{int}}[\Psi]=
\int d^2\mathbf{r}\Bigg[\beta(n_1-n_2)^2+
\frac{n^2}{2}\Bigg(\ln n-\frac{1}{2}\Bigg)\Bigg],
\label{eq:lhy_energy}
\end{equation}
and
\begin{equation}
E_{\textrm{trap}}[\Psi]=\int d^2\mathbf{r}\,V(\textbf{r})n, 
\label{eq:pot_energy}
\end{equation}
respectively. In Eq.~(\ref{eq:lhy_energy}), the first interaction term penalizes local density imbalance between the components and the second one combines the residual mean-field interaction and the LHY correction to the total-density energy. Each component norm $N_j = \int n_j d^2r$ is conserved in real time, and the total number of atoms can be calculated as $n_r\lambda^2(N_1+N_2)$.
\subsection{Homogeneous state and density control}
In the absence of an external potential, a balanced homogeneous solution is
\begin{equation}
\Psi_1=\Psi_2=\sqrt{\frac{\bar{n}}{2}}e^{-i\mu(\bar{n})\tau}, 
\label{eq:homogenous}
\end{equation}
with the background density $\bar{n}$ and chemical potential $\mu(\bar{n})=\bar{n}\ln{\bar{n}}$. The associated pressure and total-density stiffness are
\begin{equation}
P(\bar{n})=\frac{\bar{n}^2}{2}\Bigg(\ln \bar{n}+\frac{1}{2}\Bigg), 
\label{eq:pressure}
\end{equation}
\begin{equation}
g_e(\bar{n})=\left.\frac{d\mu}{dn}\right|_{\bar{n}}
=1+\ln\bar{n}.
\label{eq:g}
\end{equation}
Thus, the zero-pressure bulk density is $n_{\textrm{eq}}=e^{-1/2}\simeq0.60653$, whereas the positive compressibility requires $\bar{n}>e^{-1}$.

To explore a broader density range, we confine the mixture in a circular box,
\begin{equation}
V(\textbf{r})=
\begin{cases}
0   & r<R_b,\\
V_b & r\ge R_b,
\end{cases}
\label{eq:trap}
\end{equation}
where $r=\sqrt{x^2+y^2}$, $V_b=10$ is the height and $R_b=126$ is the radius of the potential. The background density is varied through the total norm $N=N_1+N_2$, keeping the populations balanced and the barrier parameters fixed. For a nearly uniform liquid filling the box, \(\bar n\simeq N/(\pi R_b^2)\) provides an estimate. The relevant density for the vortex theory is the measured interior plateau, away from cores and the boundary layer. An operational definition is 
\begin{equation}
\bar{n}=\frac{1}{|\textit{B}|}\int_{\textit{B}}n({\textbf{r})}d^2r,
\label{eq:n}
\end{equation}
where $\textit{B}$ is a bulk sampling region. In contrast to an unconfined self-bound droplet, the box wall supplies the pressure required to reach densities larger than the equilibrium value.
\subsection{Preparation and propagation}
We solve Eq.~\eqref{eq:dimensionless_egpe2} numerically using the split-operator Fourier pseudospectral method \cite{pseudo}. The numerical grid consists of 2048 × 2048 points in a domain
of size $(320)^2$, giving dimensionless spacings $\Delta x=\Delta y=0.156$. The time step is $dt=0.004$. Fourier propagation imposes periodicity on the computational square, so the density must be negligible at its edges. The circular wall provides the physical confinement inside that square.

An initially vortex-free state is relaxed in imaginary time at fixed, equal component norms. A vortex is imprinted in each component by multiplying the corresponding wave function by the phase factor $e^{i\Theta_j(\bm{r})}$, where
\begin{equation}
\Theta_j(\bm{r})=w_j \tan^{-1}\!\left(\frac{y-y_j}{x-x_j}\right),\quad (j=1, 2)\label{eq:orient}
\end{equation}
with $w_j=\pm1$ and $(x_j,\,y_j)$ denoting the circulation sign and the $x$-$y$ coordinates of the vortex in the $j$th component, respectively. We label the vortices such that $y_1>y_2$. The two vortices are displaced symmetrically from the center of the box along the $y$-axis by an inter-vortex distance $s$, with initial core positions $\textbf{r}_{1,2}=(0,\pm s/2)$. A short imaginary-time relaxation reduces the excitation energy introduced by imprinting.
\section{Asymptotic Intervortex Interaction}
\label{s:intervortex}
\subsection{Vortex profiles and density response}
\begin{figure}[t]
\centering
\includegraphics[width=3.4in]{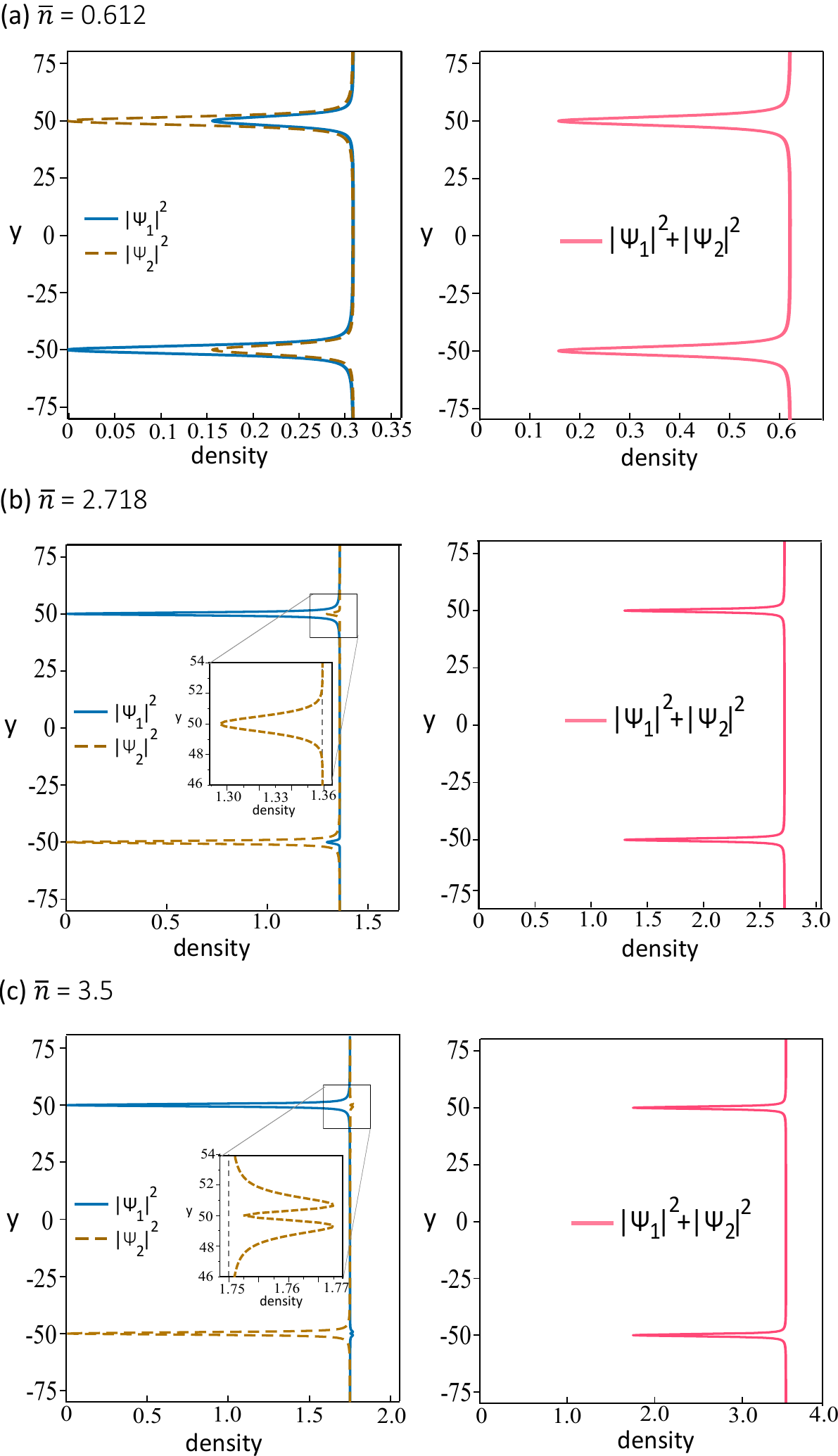}
\caption{Relaxed density profiles along $x=0$ for vortices at $(0,\pm50)$, with $\beta=1$. The left and right columns show the individual and total densities, respectively, for (a) $\bar{n}=0.612$, (b) $\bar{n}=2.718$, and (c) $\bar{n}=3.5$. The insets in (b) and (c) enlarge the density of the non-winding component near the vortex core at $(0,50)$. The non-winding component remains nonzero at the vortex center and exhibits a transition from depletion to enhancement as the background density increases. Coordinates and densities are normalized by $\lambda$ and $n_r$, respectively.}
\label{fig:1}
\end{figure}

We label a vortex in component 1 by "$(w_1,0)$" and one in component 2 by "$(0,w_2)$" with circulation direction $w_j=\pm1$. Each component has an independent U(1) phase. Around a (1,0) vortex, the phase of component 1 changes by $2\pi$ and that of component 2 does not change. The common phase $(\Theta_1+\Theta_2)/2$ therefore winds by $\pi$. For equal far-field densities, the resulting mass circulation is half that of a vortex winding both components, hence the term \textit{half-quantized vortex}. The phase winding in the occupied component itself is an integer. Figure~\ref{fig:1} shows the density profiles of the individual components and the total density. The density of the winding component vanishes at the vortex center, whereas the non-winding component retains a finite density.

For an isolated \((w_1,0)\) vortex in a uniform background, we write the stationary fields as
\begin{equation}
\Psi_1^{(w_1,0)}=\sqrt{\frac{\bar{n}}{2}}\,u(r)e^{iw_1\theta},
\qquad
\Psi_2^{(w_1,0)}=\sqrt{\frac{\bar{n}}{2}}\,v(r), 
\label{eq:ans}
\end{equation}
where $r$ and $\theta$ are polar coordinates. The real functions satisfy \(u(0)=0\), \(v'(0)=0\), and \((u,v)\rightarrow(1,1)\) as \(r\rightarrow\infty\), with the prime denotes differentiation with respect to $r$. The corresponding \((0,w_2)\) configuration is obtained by exchanging the component indices. 

By linearizing the stationary extended GP equation with $V(\textbf{r})=0$,
\begin{equation}
  \begin{aligned}
&\mu(\bar{n})\Psi_j=[-\nabla^2+n\ln n+2\beta(n_j-n_k)]\Psi_j,\quad j\neq k
\label{eq:station1}
\end{aligned}
\end{equation}
around the uniform background state and retaining only the leading density perturbations, we obtain two associated healing lengths
\begin{equation}
 \xi_+^{-2}
=2\Bar{n}(1+\ln\Bar{n})=2\Bar{n}g_{e}(\Bar{n}),
\qquad
\xi_{-}^{-2}=4\beta \Bar{n}.
\label{eq:heal}
\end{equation}
The healing lengths $\xi_+$ and $\xi-$ characterizes the recovery of total-density and relative-density perturbations, respectively. For $r\gg \textrm{max}(\xi_{+}, \xi_{-})$, the vortex profiles approach their asymptotic forms,
\begin{subequations}
\label{eq:sub}
\begin{equation}
\begin{aligned}
u(r)&=1-\frac{\xi_+^2+\xi_-^2}{2r^2}
+O(r^{-4}),
\end{aligned}
\tag{18a}
\end{equation}
\begin{equation}
\begin{aligned}
v(r)&=1-\frac{\xi_+^2-\xi_-^2}{2r^2}
+O(r^{-4}).
\end{aligned}
\tag{18b}
\end{equation}
\end{subequations}The detailed derivation of Eq.~(\ref{eq:sub}) is given in the Appendix A. In particular, the far-field perturbation of the unwound component is
\begin{equation}
\delta n_2(r)
=-\frac{\bar{n}}{2r^2}
(\xi_+^2-\xi_-^2)
+O(r^{-4}).    
\end{equation}
Thus, the far-field density response of the unwound component changes sign at \(g_e(\bar{n})=2\beta\). It corresponds to a depletion for \(g_e(\bar{n})<2\beta\) and an enhancement for \(g_e(\bar{n})>2\beta\), as illustrated in the Fig.~\ref{fig:1}. At lower density, total-density depletion is less costly than component imbalance, so both components can deplete. At higher density, preserving the total density becomes energetically more favorable than maintaining equal component densities, and the unwound component compensates for the vortex-induced depletion.
\subsection{Intervortex interaction force}
For well-separated cores at $\textbf{r}_{1,2}=(0,\pm s/2)$, we introduce the local polar coordinates \(\rho_j=|\mathbf r-\mathbf r_j|\) and an azimuthal angle \(\theta_j\) measured counterclockwise from the positive \(x\) direction about the \(j\)th vortex core (\(j=1,2\)), and use the product ansatz
\begin{subequations}
\label{eq:rescanss}
\begin{equation}
  \begin{aligned}
&\Psi_1^{(w_1,w_2)}&\simeq
\sqrt{\frac{\Bar{n}}2}\,
u(\rho_1)v(\rho_2)e^{iw_1\theta_1},
\end{aligned}
\label{eq:rescan}
\end{equation}
\begin{equation}
  \begin{aligned}
&\Psi_2^{(w_1,w_2)}&\simeq
\sqrt{\frac{\Bar{n}}2}\,
v(\rho_1)u(\rho_2)e^{iw_2\theta_2}.
\end{aligned}
\label{eq:rescan2}
\end{equation}
\end{subequations}
At fixed background chemical potential $\mu(\bar{n})$, the appropriate functional is the grand potential $\eta=E-\mu(\bar{n})N$. Subtracting the two isolated vortex contributions and adding back the uniform reference gives the pair interaction as 
\begin{equation}
\begin{aligned}
U(s/2)=\int d^2r\,
\left(\delta\mathcal K+\delta\mathcal W\right),
\label{eq:total}
\end{aligned}
\end{equation}
with \(\delta\mathcal K=\mathcal K[\Psi^{(w_1,w_2)}]-\mathcal K[\Psi^{(w_1,0)}]-\mathcal K[\Psi^{(0,w_2)}]\) and \(\delta\mathcal W=\mathcal W[\Psi^{(w_1,w_2)}]-\mathcal W[\Psi^{(w_1,0)}]-\mathcal W[\Psi^{(0,w_2)}]+\mathcal W_0\) being the differences in kinetic and local grand-potential densities, respectively. Here, \(\mathcal W_0\) denotes the local grand-potential density of the uniform, vortex-free background with \(n_1=n_2=\bar n/2\).

By substituting Eqs.~(\ref{eq:sub}) and (\ref{eq:rescanss}) in Eq.~(\ref{eq:total}), the individual self-energy contributions cancel out. The leading surviving terms
\begin{equation}
\delta\mathcal K
\simeq
-\frac{\bar{n}(\xi_+^2-\xi_-^2)}
{\rho_1^2\rho_2^2}+O(r^{-6}) ,
\label{eq:diffk}
\end{equation}
\begin{equation}
\delta\mathcal W
\simeq
\frac{\bar{n}(\xi_+^2-\xi_-^2)}
{2\rho_1^2\rho_2^2}+O(r^{-6}),
\label{eq:diffp}
\end{equation}
produce the leading interaction
\begin{equation}
U(s/2)=-\pi\Bigg[\frac{1}{g_e(\bar{n})}-
\frac{1}{2\beta}\Bigg]\frac{\ln[s/(2\xi)]}{s^2}.
\label{eq:interaction}
\end{equation}
\begin{figure}[t]
\centering
\includegraphics[width=3.4in]{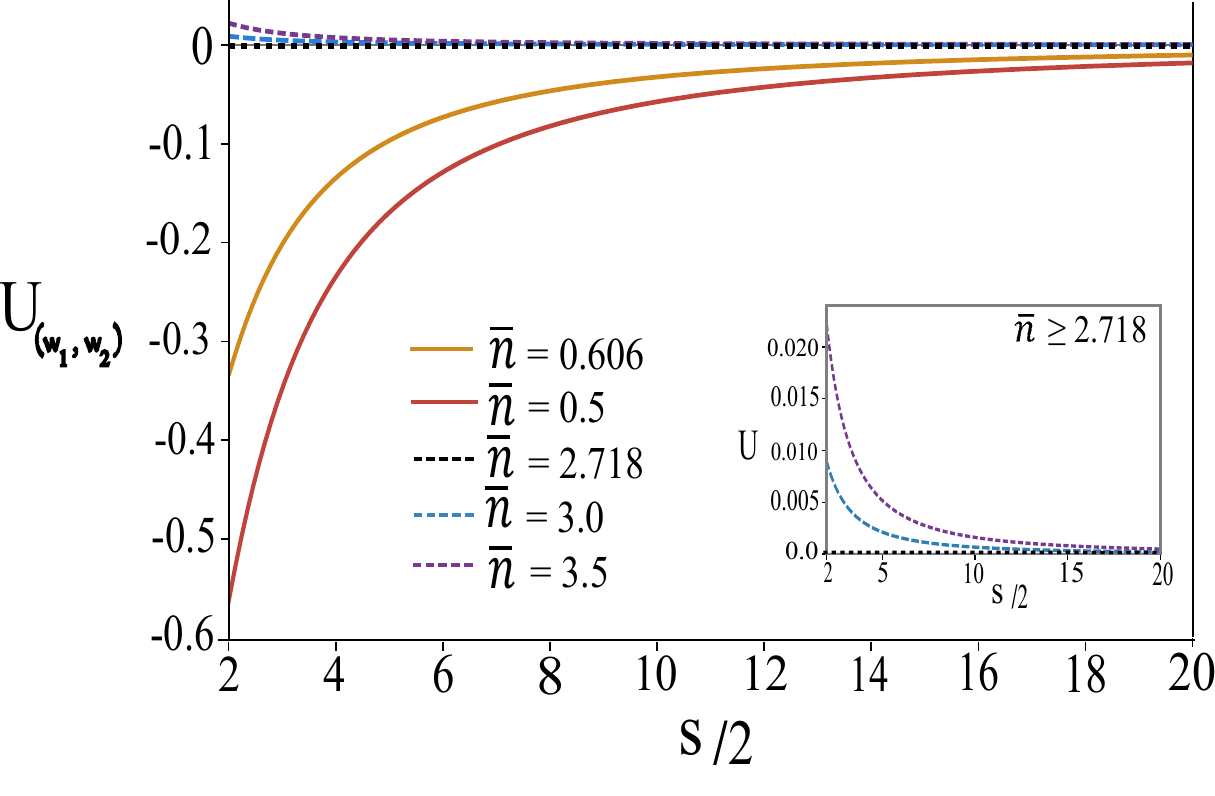}
\caption{Intervortex potential in Eq.~(\ref{eq:interaction}) as a function of half separation $s/2$ for $\beta=1$ and $\xi=\xi_{-}$. The interaction is attractive for $\bar{n}<e$, vanishes at $\bar{n}=e$, and becomes repulsive for $\bar{n}>e$. The inset enlarges the positive branch of the potential.}
\label{fig:2}
\end{figure}Here, we have used the relation \(\bar{n}(\xi_+^2-\xi_-^2)=\frac{1}{2g_e(\bar{n})}-\frac{1}{4\beta}\) and \(\xi=\xi(\bar{n},\beta)\) is an effective short-distance cutoff accounting for matching the asymptotic profiles to the nonlinear core structure, such that \(s/2\gg\xi\). For the detailed calculation of Eqs.~(\ref{eq:diffk})-(\ref{eq:interaction}), see Appendix B. The logarithmic spatial dependence also occurs in ordinary two-component BECs. The distinctive feature here is that its coefficient becomes density dependent through the LHY correction and can change sign as the background density is varied. 

The asymptotic force between the two vortices is obtained by differentiating Eq.~(\ref{eq:interaction}) with respect to the separation $s$ as 
\begin{equation}
F_s=-\frac{\partial U}{\partial s}
\simeq
-2\pi\Bigg[\frac{1}{g_e(\bar{n})}-
\frac{1}{2\beta}\Bigg]\frac{\ln[s/(2\xi)]-1/2}{s^3}.
\label{eq:force}
\end{equation}
Thus, in the asymptotic regime, the leading interaction is attractive for \(g_e(\Bar{n})<2\beta\) and repulsive for \(g_e(\Bar{n})>2\beta\). The leading interaction vanishes for \(g_e(\bar{n}_c)=2\beta\), giving the critical background density,
\begin{equation}
\bar{n}_c=e^{2\beta-1}.
\label{eq:critical}
\end{equation}
\begin{figure}[t]
\centering
\includegraphics[width=3.4in]{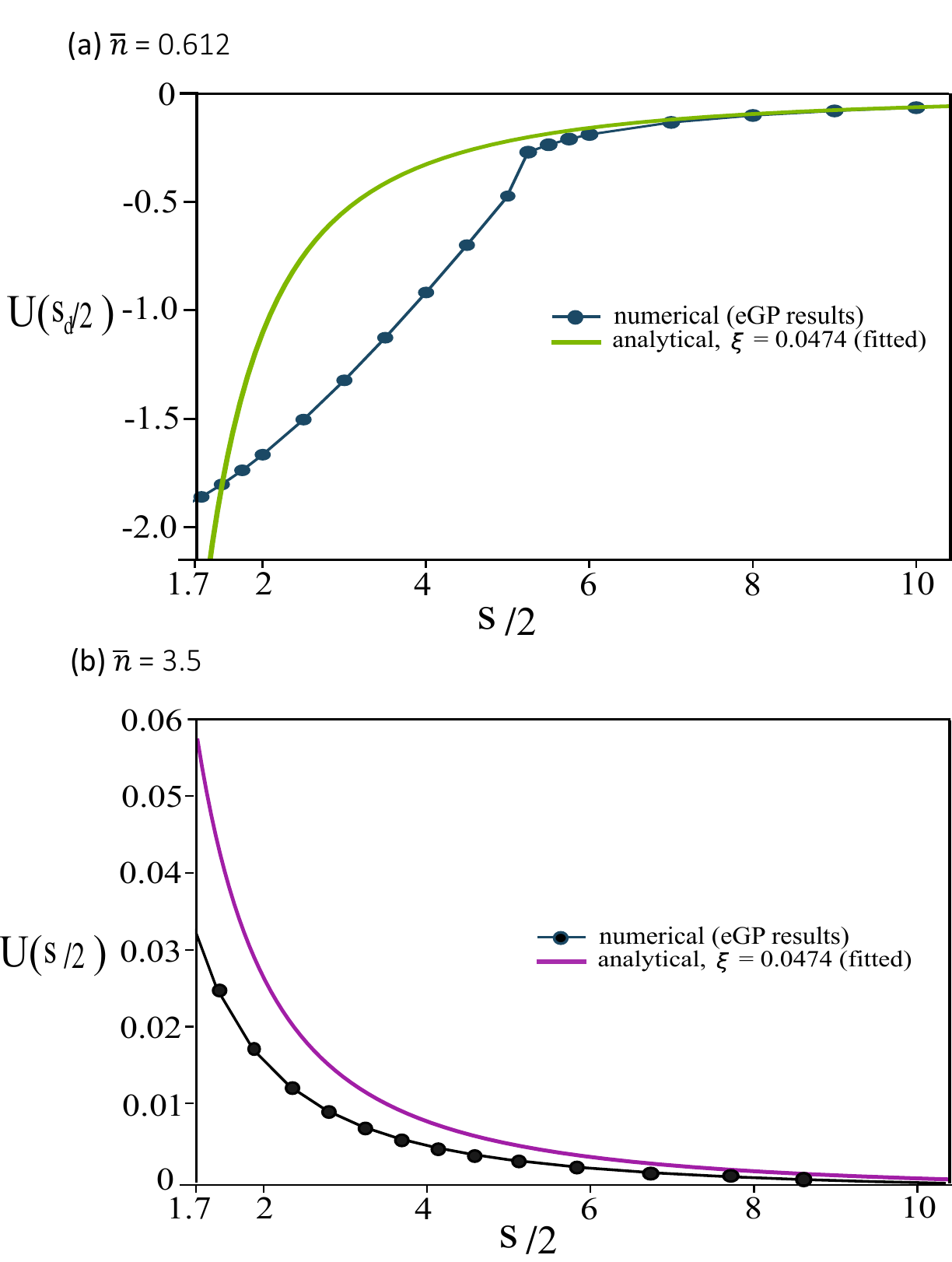}
\caption{Comparison between the asymptotic interaction potential $U(s_d/2)$ in Eq.~(\ref{eq:interaction}) and numerical results obtained by pinned-core GP minimization for $\beta=1$ and (a) $\bar{n}=0.612$ with (b) $\bar{n}=3.5$. The solid lines represent the analytical result with $\xi$ fitted to the numerical results in the asymptotic regime.}
\label{fig:3}
\end{figure}This sign reversal originates from the competition between the total-density stiffness $(g_e(\bar{n}))$ and the relative-density stiffness $(2\beta)$. At low density, where \(g_e(\bar{n})<2\beta\), total-density depletion is less costly than relative-density distortion, allowing both components to deplete around a vortex and resulting in an attractive interaction. In contrast, at high density, where \(g_e(\bar{n})>2\beta\), preserving the total density becomes energetically more favorable, and the unwound component compensates for the vortex-induced depletion, leading to repulsion. Figure~\ref{fig:2} illustrates the resulting transition from attractive to repulsive intervortex interaction.

\begin{figure*}[t]
\centering
\includegraphics[width=\textwidth]{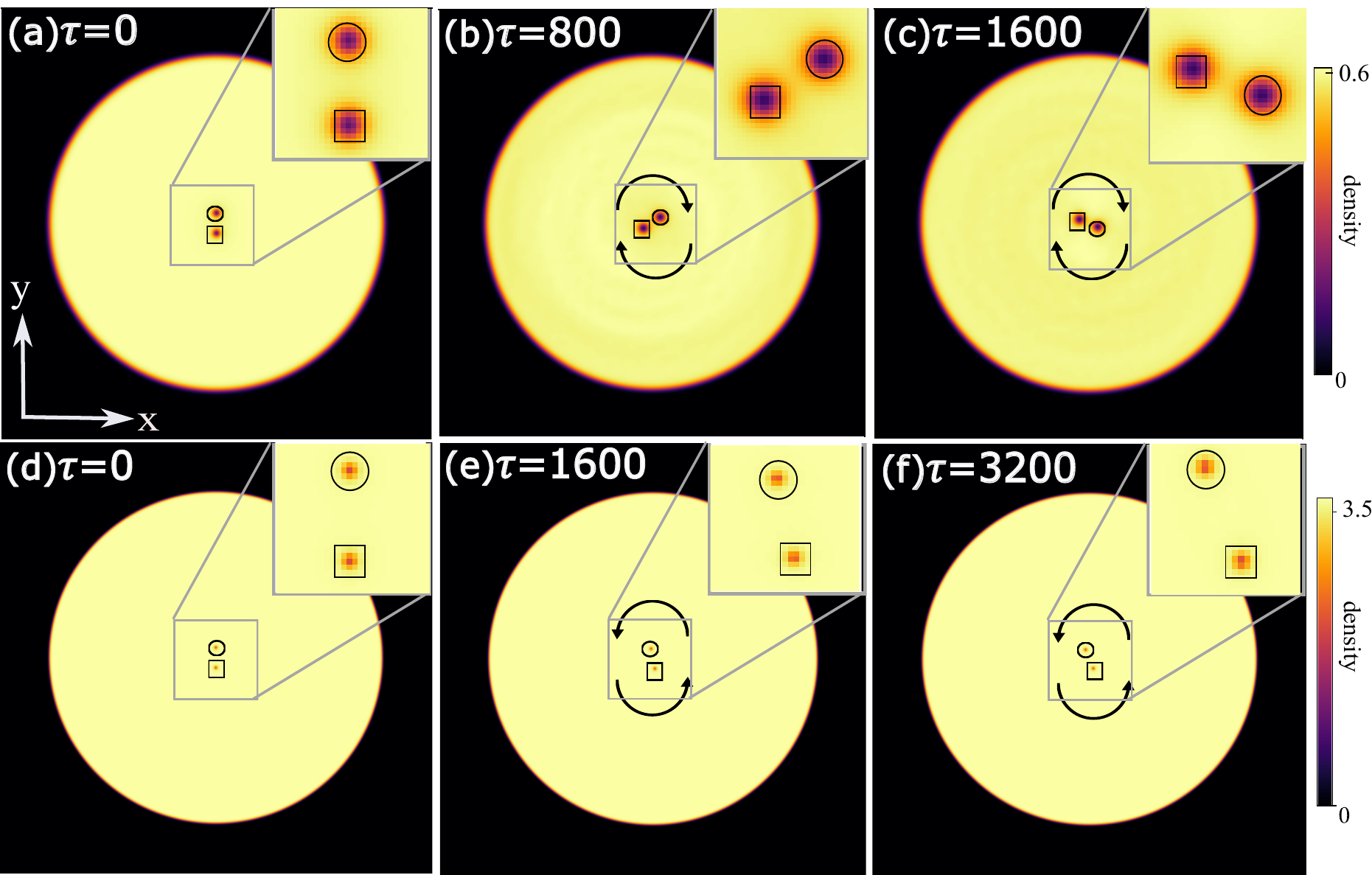}
\caption{Real-time density snapshots of $(w_1,0)$ and $(0,w_2)$ vortices having same circulation ($w_j=1$) obtained by the extended GP eq.~\eqref{eq:dimensionless_egpe2} with $\beta=1$. (a--c) show clockwise motion for $\bar{n}=0.612$ at $\tau=0, 800,$ and $1600$; panels (d--f) show counter-clockwise motion for $\bar{n}=3.5$ at $\tau=0, 1600,$ and $3200$. Circle and square represent component 1 and 2, respectively, and the insets enlarge the central region. The initial spacing is set to be $s=15$. See the Supplemental Material \cite{supp} for animations of the dynamics in (a)-(c) and (d)-(f). Other conditions are the same as those in Fig.~\ref{fig:1}.}
\label{fig:4}
\end{figure*}
It should be noted that the asymptotic interaction is independent of the circulation signs. Reversing either \(w_j\) leaves the static interaction unchanged because the two components are coupled only through densities $n_1$ and $n_2$. Therefore, all combinations of circulation signs share the same asymptotic interaction given by Eq.~(\ref{eq:interaction}).

The interaction potential is numerically estimated by constrained minimization of Eq.~(\ref{eq:dimensional_energy}), with \(\beta=1\) and \(\mu(\bar{n})=\bar{n}\ln{\bar{n}}\). For each vortex separation \(s\), we obtain the relaxed density profiles using the limited-memory Broyden-Fletcher-Goldfarb-Shanno (L-BFGS) algorithm \cite{lbfgs}. Writing the condensate fields as \(\Psi_j(\mathbf r)=A_j(\mathbf r)e^{iw_j\theta_j(\mathbf r)}\), we optimize the real amplitudes \(A_j\) while holding the phase profiles fixed to maintain the prescribed vortex positions and circulations. The algorithm uses gradients and information from recent iterations to approximate the local curvature of the grand potential, enabling efficient minimization without storing the full Hessian matrix. The numerical interaction potential is then calculated as
\[
U_{\mathrm{num}}(s)
=\eta_{12}(s)-\eta_1(\mathbf r_1)-\eta_2(\mathbf r_2)+\eta_0,
\tag{27}
\]
where \(\eta_{12}(s)\), \(\eta_j(\mathbf r_j)\), and \(\eta_0\) denote the minimized grand potentials of the vortex pair, an isolated vortex in component \(j\) at \(\mathbf r_j\), and the vortex-free background, respectively. All configurations are relaxed independently on the same computational grid at the same chemical potential and with identical boundary conditions. This subtraction removes the individual vortex contributions and restores the uniform background reference, isolating the interaction between the two vortices.

Figure~\ref{fig:3} compares the numerical intervortex interaction potential with the analytical asymptotic prediction in Eq.~(\ref{eq:interaction}) for $\bar{n}=0.612$ and $\bar{n}=3.5$. The analytical expression captures the numerical behavior in the asymptotic regime, \(5<s/2\lesssim10\). The only fitting parameter is the short-range cutoff distance $\xi$, which is determined by fitting Eq.~(\ref{eq:interaction}) to the numerical interaction potential in the asymptotic regime. At \(\bar n=0.612\) [Fig.\ref{fig:3}(a)], the numerical curve exhibits a sudden deviation from the asymptotic prediction as the separation decreases below \(s/2\simeq5\). This behavior is consistent with substantial overlap and deformation of the vortex-induced density depletions, for which the well-separated approximation breaks down. In contrast, at \(\bar n=3.5\) [Fig.\ref{fig:3}(b)], the numerical potential instead remains smooth because the non-winding component fills the density depletion at the vortex core of the winding component, reducing the total-density depletion and allowing the two vortex cores to remain distinct even at smaller separations.
\section{Reversal of Vortex Motion}
\label{s:motion}
\subsection{Magnus response to the interaction}
The density-dependent sign reversal of the intervortex interaction potential leads to a reversal of vortex motion. The motion of well-separated vortices in a locally uniform fluid can be described by the Magnus relation. Treating each vortex core as a massless point vortex located at $\mathbf{r}_j$, and neglecting any imposed background flow, the interaction force
$\mathbf{F}_j$ is balanced by the Magnus force 
\begin{equation}
\mathbf{F}_j+2\pi w_j n_{j,bg}\,\hat{\mathbf{z}}\times\frac{d\mathbf{r}_j}{d\tau}=0,
\label{eq:bal}
\end{equation}
where $n_{j,bg}=\bar{n}/2$ is the background density of the $j$th component \cite{orb1}. Equation~(\ref{eq:bal}) then gives
\begin{equation}
\frac{d\mathbf{r}_{j}}{d\tau}=\frac{\hat{\mathbf{z}}\times\mathbf{F}_j}{2\pi w_j n_{j,bg}}, \label{eq:mag}
\end{equation}
which indicates that the vortex velocity is perpendicular to the force acting on it. The direction is obtained by rotating $\mathbf{F}_j$ by $\pm\pi/2$ for
$w_j=\pm1$. For two vortices with $w_j=1$ at $(0,\pm s/2)$, $\mathbf{F}_1=F_s\,\hat{\mathbf{y}}$ and $\mathbf{F}_2=-F_s\,\hat{\mathbf{y}}$, and therefore attraction drives the upper and lower cores towards $+x$ and $-x$ directions, respectively, resulting in clockwise motion. Repulsion reverses these velocities and gives counter-clockwise motion. A radial interaction force therefore produces transverse orbital motion; attraction does not necessarily imply that the vortex cores move towards one another in this non-dissipative point-vortex limit.

Taking counter-clockwise angular velocity as positive, Eqs.~(\ref{eq:force}) and (\ref{eq:mag}) give
\begin{equation}
\Omega_j=-\frac{4}{\bar{n}}\Bigg[\frac{1}{g_e(\bar{n})}-
\frac{1}{2\beta}\Bigg]\frac{\ln[s/(2\xi)]-1/2}{s^4},
\label{eq:angular}
\end{equation}
which implies a reversal of the orbital motion at $g_e(\bar{n})=2\beta$. Opposite circulation signs leave the static potential unchanged but reverse the Magnus response, predicting the translation of the vortex pair. The direction of the translation also changes its sign at $g_e(\bar{n})=2\beta$. The numerical results presented below are restricted to the case of equal positive circulations, $w_1=w_2=1$.
\begin{figure}[t]
\centering
\includegraphics[width=3.4in]{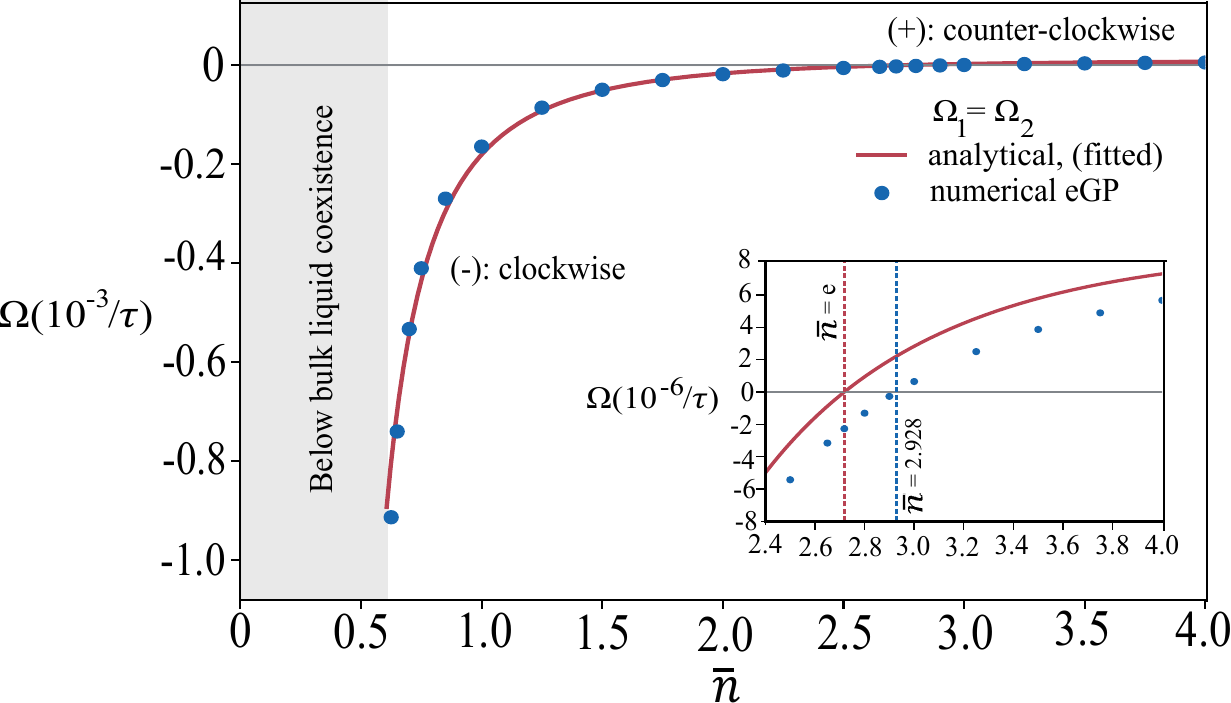}
\caption{Initial angular velocity versus total background density $\bar{n}$ with $s\simeq15$ and $\beta=1$. The line and dots represent the analytical cutoff fit and numerical extended GP results, respectively. Negative and positive values denote clockwise and counter-clockwise motion. The inset resolves the difference between the analytical zero at $\bar{n}=e$ and the numerical crossing labeled $\bar{n}=2.928$.}
\label{fig:5}
\end{figure}
\subsection{Numerical dynamics and the reversal density}
Figure~\ref{fig:4} shows the real-time evolution obtained by solving the extended GP equation numerically, where $(w_1,0)$ and $(0,w_2)$ vortices with $w_j=1$ are initially located at $(0, \pm7.5)$. At the lower background density, \(\bar n=0.612\), the vortex cores undergo clockwise orbital motion [Figs.~\ref{fig:4}(a)–\ref{fig:4}(c)]. At the higher density, \(\bar n=3.5\), the direction of rotation reverses, and the vortices exhibit substantially slower counterclockwise motion [Figs. \ref{fig:4}(d)–\ref{fig:4}(f)]. These observations are consistent with the Magnus response described in Sec.~\ref{s:motion} A, which predicts clockwise and counterclockwise motion for the attractive and repulsive intervortex interactions, respectively.

Figure~\ref{fig:5} quantifies the density dependence of the initial angular velocity, extracted from the short-time vortex trajectories. Counterclockwise rotation is taken as positive, so that \(\Omega<0\) and \(\Omega>0\) correspond to clockwise and counterclockwise motion, respectively. The analytical prediction captures the strong reduction in the magnitude of the clockwise angular velocity as the background density approaches the reversal point. The inset enlarges the small angular velocities near and above the crossing. On the higher-density side, the numerical angular velocity becomes positive and increases over the range shown, while remaining substantially smaller in magnitude than the clockwise angular velocity at low density.

The numerical reversal occurs near \(\bar n\simeq2.928\), whereas the leading asymptotic theory predicts a vanishing interaction coefficient at \(\bar n_c=e\). Near \(\bar n_c\), the leading contribution is suppressed, making the dynamics more sensitive to subleading interaction terms, finite vortex separation, and possible confinement-induced motion. Such effects may shift the numerical reversal point away from the asymptotic prediction.
\section{Conclusions}
\label{sec:conc}
We have investigated the interaction and dynamics of vortices belonging to different components in a 2D Bose quantum liquid. Using the logarithmic extended-GP model, we derived the density-dependent asymptotic interaction between vortices. Its leading coefficient is set by the difference between the inverse total-density and relative-density stiffnesses. As the background density increases, the sign of the leading coefficient changes at a critical density $\bar{n}_c=e^{2\beta-1}$, leading to a reversal of the intervortex interaction from attractive to repulsive. Real-time simulations revealed a corresponding reversal of vortex motion. For two vortices with equal positive circulation, clockwise orbital motion at low density changes to counterclockwise motion at high density. This behavior was explained by the Magnus response to the density-dependent interaction force.

These results demonstrate that interactions between topological defects and their motion can be controlled by the equation of state in multicomponent quantum liquids. Future work could investigate the short-range interaction and the coupling of vortex
motion to compressible and surface excitations, which become increasingly important near the interaction-reversal density.
\section*{ACKNOWLEDGMENT}
We acknowledge support from the National Science and Technology Council (NSTC), Taiwan, under the Grants No. NSTC-115-2112-M-001-035-MY3 and No. NSTC-115-2119-M-001-009, and from Academia Sinica under Grant AS-CDA-113-M04 and for support from TG 1.2 of NCTS at NTU. H.S. was supported by JSPS KAKENHI Grants No. JP23K03276 and No. JP26K00638. 
\section*{DATA AVAILABILITY}
The data supporting the findings of this article are not publicly available at the time of publication due to technical limitations. The data are available from the authors upon reasonable request.

\section*{Appendix A: Far-field vortex profiles}
Substituting Eq.~(\ref{eq:ans}) into Eq.~(\ref{eq:station1}), gives 
\begin{subequations}
\begin{equation*}
\begin{aligned}
-u''-\frac{u'}r+\frac{u}{r^2}
+\left[
n\ln n-\mu(\bar n)
+\beta\bar n(u^2-v^2)
\right]u=0,
\end{aligned}
\tag{A1}
\end{equation*}
\begin{equation*}
\begin{aligned}
&&&&&-v''-\frac{v'}r
+\left[
n\ln n-\mu(\bar n)
-\beta\bar n(u^2-v^2)
\right]v=0, 
\end{aligned}
\tag{A2}
\end{equation*}
\end{subequations}
where
\(n=\bar n(u^2+v^2)/2\) and
\(\mu(\bar n)=\bar n\ln\bar n\).
Setting \(u=1+a_1\), \(v=1+a_2\), and introducing \(a_\pm=a_1\pm a_2\), the linearized equations decouple into
\begin{subequations}
\begin{equation*}
\begin{aligned}
\left[
-\frac{d^2}{dr^2}-\frac1r\frac{d}{dr}
+2\bar n\,g_{\rm e}(\bar n)
\right]a_+
&=-\frac1{r^2}+O(r^{-4}),
\end{aligned}
\tag{A3}
\end{equation*}
\begin{equation*}
\begin{aligned}
\left[
-\frac{d^2}{dr^2}-\frac1r\frac{d}{dr}
+4\beta\bar n
\right]a_-
&=-\frac1{r^2}+O(r^{-4}).
\end{aligned}
\tag{A4}
\end{equation*}
\end{subequations}
Since \(\nabla^2r^{-2}=4r^{-4}\) away from the origin, the leading terms are \(
a_\pm=-\xi_\pm^2/r^2+O(r^{-4})\). Taking their sum and difference gives Eq.~(\ref{eq:sub}). Since the centrifugal term is independent of the sign of the circulation, reversing the circulation leaves the density tails unchanged.
\section*{Appendix B: Leading logarithmic interaction}
We define \(q=n_1-n_2\), \(\delta n=n-\bar n\), and the local grand-potential density 
\[
\mathcal W(n,q)
=\frac{1}{2}n^2\Bigg(\ln{n}-\frac{1}{2}\Bigg)+\beta q^2-\mu(\bar n)n.
\tag{B1}
\]
Expanding about $(\bar{n}, 0)$ gives
\[
\mathcal W(n,q)-\mathcal W(\bar n,0)
=\frac12g_{\rm e}(\bar n)(\delta n)^2
+\beta q^2+O((\delta n)^3).
\tag{B2}
\]
The linear term cancels because the subtracted chemical potential is that of the chosen background. The isolated vortex tails are
\begin{subequations}
\begin{equation*}
\begin{aligned}
\delta n^{(1)}&=-\frac{\bar n\xi_+^2}{\rho_1^2},
&
q^{(1)}&=-\frac{\bar n\xi_-^2}{\rho_1^2},
\end{aligned}
\tag{B3}
\end{equation*}
\begin{equation*}
\begin{aligned}
\delta n^{(2)}&=-\frac{\bar n\xi_+^2}{\rho_2^2},
&
q^{(2)}&=+\frac{\bar n\xi_-^2}{\rho_2^2}.
\end{aligned}
\tag{B4}
\end{equation*}
\end{subequations}
The quadratic cross term is therefore
\begin{subequations}
\begin{equation*}
\begin{aligned}
\delta\mathcal W
&=g_{\rm e}(\bar n)
\,\delta n^{(1)}\delta n^{(2)}
+2\beta q^{(1)}q^{(2)}
\end{aligned}
\end{equation*}
\begin{equation*}
\begin{aligned}
&=\frac{\bar n(\xi_+^2-\xi_-^2)}
{2\rho_1^2\rho_2^2}.
\end{aligned}
\tag{B5}
\end{equation*}
\end{subequations}
The corresponding phase-gradient contribution is
\[
\begin{aligned}
\delta\mathcal K
=\frac{\bar n}{2}\Bigg[
\frac{u^2(\rho_1)[v^2(\rho_2)-1]}{\rho_1^2}
+\frac{u^2(\rho_2)[v^2(\rho_1)-1]}{\rho_2^2}
\Bigg].
\end{aligned}
\]
\[
\simeq
-\frac{\bar n(\xi_+^2-\xi_-^2)}
{\rho_1^2\rho_2^2}.
\tag{B6}
\]
Amplitude-gradient terms do not change the leading logarithmic coefficient, although core regions can contribute to non-logarithmic remainder.

To evaluate the coefficient, consider 
\[
I=
\int
\frac{d^2r}
{\rho^2_1\rho^2_2}.
\]
We exclude disks of a matching radius $a$ around each core and choose a fixed $0<c<1/2$. In the annulus $a\ll\rho_1<cs$, the other core distance is of order $s$, and the logarithmically divergent contribution is 
\[
I= \frac{2\pi}{s^2}\ln{\bigg(\frac{cs}{a}\bigg)}+O(s^{-2}).
\tag{B7}
\]
The second annulus contributes the same amount. After rescaling $\textbf{r}=s\textbf{x}$, the remaining region gives a finite constant times $s^{-2}$. Thus
\[
I= \frac{4\pi}{s^2}\ln{\bigg(\frac{s}{a}\bigg)}+O(s^{-2}).
\tag{B8}
\]
Matching to the nonlinear cores determines the constant term and the effective cutoff convention. Combining this result with Eqs.(B5) and (B6), and using
\[
\begin{aligned}
\bar{n}(\xi_+^2-\xi_-^2)=\frac{1}{2g_e(\bar{n})}-\frac{1}{4\beta},
\end{aligned}
\tag{B9}
\]
gives Eq.~(\ref{eq:interaction}). This argument fixes the logarithmic coefficient but does not promote the cutoff-dependent constant to an exact asymptotic result.
\bibliographystyle{plain}

\end{document}